# Investigations of Metal-Organic structures for applications in Organic Spin Valve devices

Sonia Kaushik, Manisha Priyadarsini, Avinash G. Khanderao, Dileep Kumar[a)]

UGC-DAE Consortium for Scientific Research, Khandwa Road, Indore-452001, India
[a)]dkumar@csr.res.in

## Abstract

The integration of the immense potential of spintronics and the advantages of organic materials led to the emergence of organic spintronics. Organic spintronics has drawn the interest of the science community due to various applications in spin-valve devices. But to date, an efficient room-temperature Organic Spin Valve device has not been experimentally realized due to the complicated spin transport at the metal-organic interfaces. These studies are always challenging due to the complicated spin-polarized charge transfer at the metal-organic interfaces. The present study focuses on a comprehensive understanding of the interfacial properties that are essential for advancing device performance and functionality. The Ferromagnetic metals and half-metallic electrodes such as Co, $Co_2FeAl$, etc., and fullerene ($C_{60}$) bilayer samples are prepared and studied via different structural and magnetic characterizations. Due to the mechanical softness of $C_{60}$, deep penetration of ferromagnetic metal atoms is observed inside the $C_{60}$ film. In-situ MOKE measurements reveal the origin of the 23 Å thick magnetic dead layer at the interface, which is attributed to the diffused ferromagnetic clusters exhibiting superparamagnetic behavior. In contrast to the inorganic substrates, magnetic anisotropy tends to develop at 40 Å thick Co film deposited on $C_{60}$ which enhances with increasing thickness. The XRD measurements confirm the presence of in-plane compressive strain and texturing along the hcp (002) phase in the Co film. The anomaly observed in the hard axis of magnetization is due to high dispersion in the local magnetic anisotropy. The variation of the magnetic anisotropy axis is observed in the $Co_2FeAl_{wedge}$ deposited in the proximity of $C_{60}$. These findings provide valuable insights into the complex interplay between ferromagnetic materials and organic semiconductors offering potential avenues for tailoring magnetoresistance effects and fundamental understanding of organic spintronic devices.

**Keywords:** Metal-organic interface; Organic Spintronics; magneto-optic Kerr effect

## 1. Introduction

The exploration of possibilities introduced by organic semiconductors is a rapidly growing field. Their application in developing thin-film transistors, organic solar cells, and organic LEDs [1–3] represents one of the most significant accomplishments in this research area. In addition to facilitating charge transport, organic semiconductors have been investigated for their potential to mediate spin transport [4–6]. However, achieving the goal of organic spintronics—specifically, designing an efficient organic spin-valve (OSV) device—remains challenging due to the complex transport phenomena at the interfaces. OSV devices typically consist of a three-layer structure, where organic semiconductors such as fullerene (C60), pentacene, and tris(8-hydroxyquinolinato)aluminum (Alq3) are sandwiched between ferromagnetic layers. Because organic semiconductors are composed of low atomic number elements, they exhibit weak spin–orbit coupling, reduced hyperfine interactions, extended spin relaxation times, and offer thermal and mechanical flexibility, all of which make them promising candidates for spintronic applications [7–9].

In the design of an effective OSV, the properties of the interfaces are critical [10,11]. In 2002, Dediu et al. first demonstrated that the organic molecule sexithienyl (T6) could serve as a spin transport channel in a lateral structure consisting of LSMO (La$_{2/3}$Sr$_{1/3}$MnO$_3$)/T6/LSMO [12]. Similarly, Xiong et al. reported the first observation of the spin-valve effect in vertical OSV devices, where tris(8-hydroxyquinolinato)aluminum (Alq3) was placed between LSMO and cobalt (Co) layers [13]. Recent investigations have shown that at the interface between a ferromagnetic metal and C60, hybridization between the metal's 3dz² orbitals and the carbon's 2pz orbitals produces unexpected magnetic behavior in both materials [14–18].

Most studies in the literature address the ideal case where an organic layer is deposited on a ferromagnetic surface, while the reverse configuration—depositing a ferromagnet onto an organic layer—is less frequently explored. For example, Wang et al. reported a cobalt penetration depth of 125 Å when deposited onto Alq3 [19]. This process results in a magnetic dead layer several nanometers thick, with a rough Co/Alq3 interface attributed to the oxidation of the cobalt surface through the organic layer. Significant efforts have been made to enhance interface quality and control the penetration of ferromagnetic electrodes, such as by introducing a tunnel barrier between the organic semiconductor and the top electrode or by employing buffer-assisted growth techniques [20–23]. Nevertheless, even with improved organic film quality, the penetration of top ferromagnetic electrodes remains difficult to regulate [24]. Researchers are also studying magnetic properties of the electrodes— including magnetic anisotropy, the formation of magnetic dead layers, and coercivity (HC) [25–27]— with the aim of correlating these characteristics to the evolving structure and morphology during film growth.

Magnetic anisotropy is a key characteristic in FM-OSC based structures, fundamentally influencing the performance and operation of organic spin valves and defining their magnetic functionality [28,29]. Recent studies have revealed that an in-plane uniaxial magnetic anisotropy (UMA)—in which spins predominantly align within the film plane—can arise from the diffused interfacial morphology found in metal-organic structures [30,31]. This preferential in-plane spin alignment is rather unexpected in FM-OM systems. In contrast, epitaxial inorganic magnetic thin films typically exhibit UMA when deposited on single-crystal substrates such as Ag (001), Cu (100), MgO (100), and SrTiO$_3$ (001) [32–34]. In these inorganic films, an ordered crystallographic structure facilitates spin–orbit coupling (SOC), which in turn establishes a preferred magnetization direction relative to the crystal lattice—a phenomenon known as magneto-crystalline anisotropy (MCA).

However, metallic thin films grown on organic materials tend to be amorphous or polycrystalline due to lattice mismatches and interfacial diffusion. As a result, such films lack long-range structural order [35], making it difficult to achieve MCA or UMA from randomly oriented grains. In contrast, the origin of UMA in organic bilayer structures is linked to the film's preparation conditions and surface roughness. For instance, I. Bergenti and colleagues deposited Co thin films onto Tris(8-hydroxyquinoline) aluminum (Alq$_3$) layers using RF sputtering, attributing the observed UMA in the Co film to the morphology of the underlying Alq$_3$ layer [30]. Similarly, V. Kalappattil et al. examined the magnetic anisotropy of the bottom electrode in an LSMO/Alq$_3$/Co spin valve structure, where substrate-induced compressive strain enhanced the magnetic anisotropy of the LSMO layer [36]. Despite the significant role of UMA, its fundamental origin in organic-ferromagnetic systems is not yet fully understood. Although previous research has correlated the morphology of the organic layer with the emergence of magnetic anisotropy, the mechanism by which an isotropic surface roughness induces such anisotropy in metal-organic systems remains unclear [31]. This underscores the need for a careful and systematic investigation into the development of magnetic anisotropy in metal-organic bilayer structures.

In the present work, ferromagnetic thin films deposited on C$_{60}$ films are studied in situ as well as ex-situ to understand the evolution of magnetic properties and to corroborate it with the structure and morphology that evolved during the film growth. Structural evolution during film growth is examined using the reflection high energy electron diffraction (RHEED) technique, whereas in-situ magneto-optic Kerr effect (MOKE) measurements have been used to study the evolution of the magnetic properties. The understanding of the magnetic anisotropy in Co/C$_{60}$ bilayer samples with thicknesses of the Co layer from 10 Å to 200 Å is done using ex-situ MOKE and X-ray diffraction (XRD) measurements to correlate magnetic anisotropy with the structure, and morphology evolved during the film growth. The deposition of $Co_2FeAl_{wedge}$ in the proximity of $C_{60}$ confirms the presence of magnetic dead layer along with the variation in anisotropy axis. The unique behaviour of rotation of the easy axis of magnetism is further correlated with the varying stress associated with the increasing thickness of the $Co_2FeAl$ layer along the sample. This approach enables us to investigate the fundamental mechanism underlying the development of magnetic anisotropy in metal-organic thin films as thickness increases. Gaining a comprehensive understanding of magnetic anisotropy in metal-organic thin films and developing effective control over it will empower researchers to tailor film properties and optimize their functionality. This knowledge and control hold significant potential for applications in organic spintronic devices, magnetic sensors, and data storage media.

## 2. Materials and Methods

### Sample Preparation

A thin layer of $C_{60}$ was prepared on a Silicon (Si) substrate via sublimation of commercially available $C_{60}$ powder at a rate of 0.2 Å/sec using a self-designed PID-controlled thermal evaporator in a high vacuum chamber (base pressure of $10^{-7}$ mbar. Ferromagnetic metals and half-metallic electrodes, such as Co, $Co_2FeAl$, etc., are deposited using e-beam evaporation in an Ultra High Vacuum chamber (base pressure ~$10^{-10}$ mbar). Magnetic anisotropy of Co films on $C_{60}$ layer, A series of four Co/$C_{60}$ bilayers were prepared by depositing with Co films of nominal thicknesses 10 Å, 20 Å, 40 Å and 200 Å on around 700 Å thick $C_{60}$ layer. Further, in the manuscript, these samples will be denoted as S1, S2, S3 and S4, respectively. Metallic layers were prepared using High-power impulse magnetron sputtering (HiPIMS) under a base pressure of $1.33 \times 10^{-7}$ mbar. The thin capping layer of Aluminium (Al) having a thickness of 20 Å was deposited to prevent the oxidation of Co film.

### In-situ Characterizations

Co thickness growth on $C_{60}$ film was monitored online using a pre-calibrated quartz crystal monitor. He-Ne laser (wavelength =632.8 nm) was used for in-situ MOKE measurements in longitudinal geometry. A dc current of 5 mA was applied across the sample's contact pads made of Au, and the film's resistance was calculated by measuring the voltage drop across the film. The structural and magnetic properties during the Co film growth, in-situ RHEED, and MOKE measurements were carried out at the different stages during the film growth.

### Ex-situ Characterizations

Before using $C_{60}$ as a template for Co growth, the thickness, uniformity, and structure of $C_{60}$ films were characterized using X-ray reflectivity (XRR) and Raman spectroscopy techniques. The thickness and structure of the films were obtained using XRR carried out using a Bruker © D8 diffractometer using x-rays of energy 8.047 keV. Synchrotron-based XRD measurements were carried out at BL02 beamline, RRCAT, Indore in out-of-plane (OP-XRD) and in-plane (IP-XRD) geometry at energy 15 keV. For the samples' topographical analysis and surface roughness measurement, atomic force microscopy (AFM) measurement in tapping mode was performed at room temperature using Bruker's Bioscope Resolve system. The silicon cantilever with nominal spring constant of 50 N/m and resonant frequency around 290 kHz was used for imaging.

Surface magnetic measurements in longitudinal geometry were carried out using the MOKE microscope of M/s Evico Magnetics, Germany. Broadband white LED light source was used in the MOKE microscopy system. The hysteresis loops were measured with varying azimuthal angle ϕ to study the magnetic anisotropy in the sample. The change of the in-plane azimuthal angle in the film plane was controlled by a rotational feedthrough with its axis perpendicular to the film plane.

## 3. Results and discussion

## 3.1 Characterization of $C_{60}$ Film

Prior to the deposition of Cobalt film, a thin layer of $C_{60}$ was deposited. Fig. 1(a) gives the XRR pattern of the as-deposited $C_{60}$ film on Silicon substrate. Sharp periodic oscillations in the XRR pattern indicate the uniform and good quality of the film. XRR pattern is fitted using Parratt formalism [37] to get the thickness, roughness, and electron density of the $C_{60}$ thin film. The fitting of the experimental XRR data is plotted with the continuous line in the same Figure, whereas the extracted scattering length density (proportional to the electron density) depth profile of the thin film is shown in Fig. 1(b). The thickness of $C_{60}$ film is found to be 1006 Å with 18 Å surface roughness. Electron density, calculated based on the SLD profile, is found to be 34% less than the calculated theoretical value of the bulk $C_{60}$ material. It is mainly due to the formation of the voids and the gaps between the $C_{60}$ molecules.

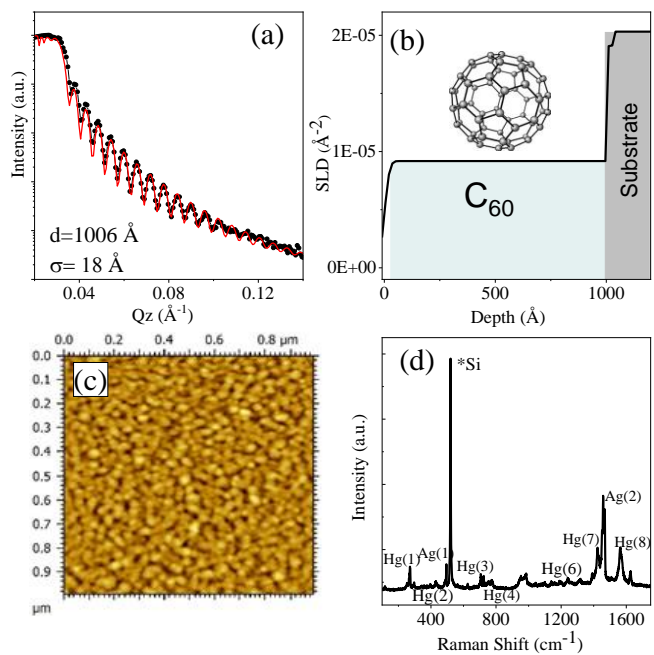

**Fig.1.** (a) XRR plot of the $C_{60}$ film on Si substrate where symbols are the experimental data points, and the continuous line shows the fitted data using Parratt formalism. (b) shows the electron scattering length density profile of the fullerene film. (c) AFM micrograph of $C_{60}$ thin film on Si substrate at room temperature. (d) Raman Spectra of $C_{60}$ film on Si Substrate.

AFM image of the film is given in Fig. 1(c). AFM image clearly shows the uniform and symmetric surface morphology of the $C_{60}$ film. The root mean square roughness of the $C_{60}$ film obtained from AFM measurements corroborates with the one obtained from the XRR measurement. To further characterize the $C_{60}$ film, Raman spectroscopy was done, and the main allowed Raman features are shown in Fig. 1(d). All the corresponding modes obtained in the spectra agree well with the literature [38,39].

## 3.2 In-situ Investigation of Co film on $C_{60}$

### 3.2.1 Resistivity measurements

To further study the magnetism at the interface of Co-$C_{60}$ and to correlate it with the intermixing and metal atom penetration; Co films, starting from a fraction of nanometer to tens of nanometers, were grown on $C_{60}$ film in uniform deposition conditions and characterized simultaneously using in-situ MOKE and transport measurements. Fig. 2(a) shows that two side Au contact pads were deposited prior to the start of Co evaporation for transport measurements. Fig. 2(b) gives the thickness dependence of normalized resistance ($R_s$) performed simultaneously with MOKE measurements. The thickness-dependent transport studies for Si/Co and Si/C$_{60}$/Co samples were carried out in two separate sets of experiments in the same UHV chamber. Therefore, the normalized resistance data is presented in Fig. 2 (b) to compare the "resistance vs thickness" trends in both samples.

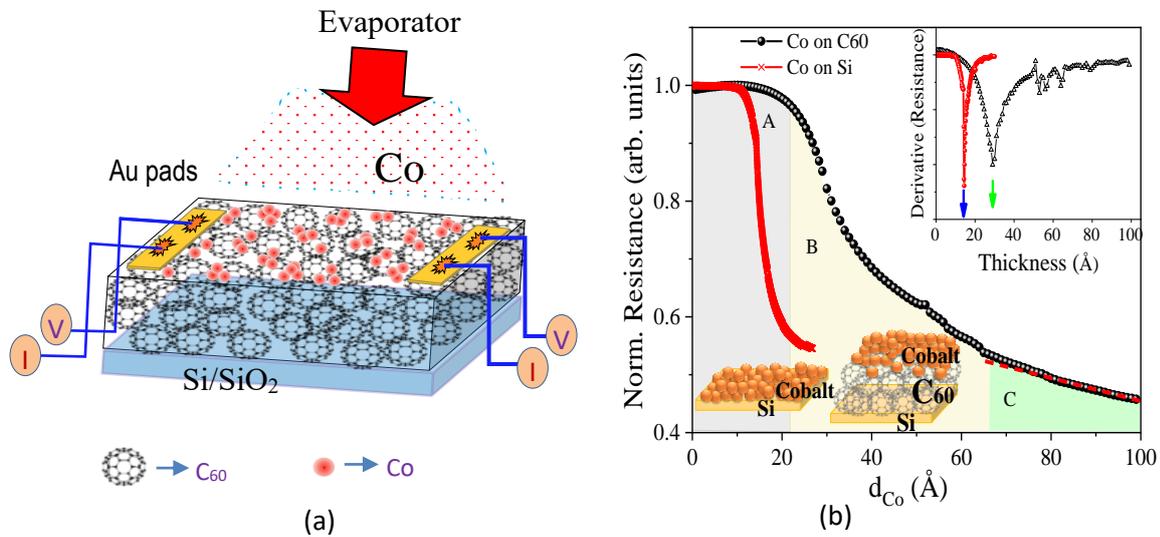

**Fig.2.** (a) Schematic illustration of Co deposition on C$_{60}$ sample (e$^-$ beam evaporation) (b) Normalized Resistance v/s thickness of Co film. Inset shows the derivative of resistance v/s thickness of Co.

The growth of Co on C$_{60}$ can be understood by dividing the resistance curve into three regions: A, B, and C. It can be seen that resistance remains constant up to the thickness of 14 Å in the region 'A', and with the further increase in the thickness, a sudden fall in the resistance in the thickness range of 20 - 50 Å can clearly be seen in the region 'B'. Beyond this thickness, a monotonous decrease in the resistance can be seen in the region 'C', which agrees with the formation of a continuous layer of Cobalt on C$_{60}$. It may be noted that C$_{60}$ is a semiconductor and possesses resistance tens of kilo-ohms [40,41]. Therefore, variation in the resistance with the increasing thickness of the Co film is mainly related to the Co layer. Resistance v/s thickness plot can be used to understand the growth of Co in different stages at increasing thickness. In Region A, constant resistance can be understood in terms of finite or small-sized islands. These islands are well separated from each other. Therefore, the possibility of an electron hopping from one island to another is negligible due to the isolated island boundaries, which work as potential barriers to electron conductivity. In this thickness range, currents conduct through the C$_{60}$ layer and the substrate. As the Co film thickness increases beyond 20 Å, the size of the islands increases and the possibilities of electronic conduction through quantum mechanical tunnelling between adjacent islands and thermionic emission increase due to the decreasing island boundaries [42]. In fact, the slight decrease in resistance at the end of region A or the starting of region B may be attributed to the quantum mechanical tunnelling between the nearest neighbouring Co islands. The drastic decrease of resistance in region B with a further increase in the Co thickness beyond ~25 Å is mainly due to the connecting path between the islands, which forms a percolation path for conducting electrons. At this stage, reduction in the separation of islands forms network-like structure and therefore, resistance decreases drastically [43]. The comparatively slower decrease in the second part of region B is due to the classical size effect. When the film thickness becomes comparable to the mean free path of the conduction electrons, the scattering of the conduction electron at the film's surface limits the conductivity of the electrons. The surface and interface scattering in region C will be less significant; therefore, the resistance decreases monotonously in this region. A similar effect has also been observed by Fuchs and Sondheimer, who showed that conductivity loss could be attributed to the scattering of the conduction electrons at the film's surface [44].

A separate sample has also been grown using e-beam evaporation in the same system to compare with Co on Si. The normalized resistance v/s thickness dependence curve is shown in Fig.. 2(b). Co film develops via islands growth, but in comparison to the growth of Co on C$_{60}$, coalescence into a continuous film takes place at around 20 Å thickness. Moreover, the percolation threshold is sharp and takes place between the thickness range of 11 to 19 Å, while on $C_{60}$, percolation lies in the thickness range of 23 to 40 Å, which is shown in the derivative of transport curves in the inset of Fig.. 2(b). Arun et al. studied the growth of e-beam evaporated Co films on Si (001)/SiO$_2$ substrate using a real-time multi-beam optical stress sensor (MOSS) and in-situ MOKE measurements. The film also grows via Volmer–Weber mechanism, where islands grow larger to impinge with other islands. The development of tensile stress in the film is associated with the island coalescence process. The film exhibits a well-defined UMA, which varies synchronously with internal tensile stress developed during growth [45]

### 3.2.2 RHEED Measurements

To investigate the structure of cobalt film during growth, in-situ RHEED measurements were performed at different thicknesses of Co film. Some representative RHEED images, collected after deposition of 12 Å, 41 Å, 71 Å, and 100 Å, are presented in Fig.. 3. A set of concentric rings (Debye rings) corresponding to different diffraction planes of Co film were observed when the sample was illuminated with the electron beam. The observed half-ring pattern in all RHEED images confirms the polycrystalline nature of the Co films. It may be noted that the underneath Si substrate is covered with natural Si oxide layer (SiO$_2$). Therefore, the substrate is not expected to participate in Co electron diffraction due to its amorphous nature.

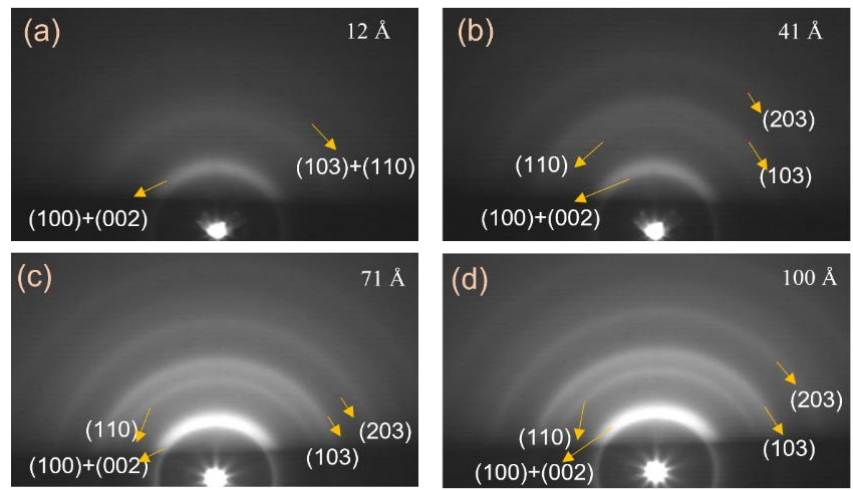

**Fig. 3.** RHEED patterns of Cobalt film of the thickness (a) 12 Å, (b) 41 Å, (c) 71 Å, (d) 100 Å.

In order to visualize Co diffraction peaks, pixel intensities are plotted with increasing 2θ angles (θ=0 corresponds to the direct electron beam spot at the centre of the images). Intensity v/s 2θ plots for all four images are shown in Fig. 4(a). Peaks are identified for their corresponding Co hcp planes and are mentioned in the same Figure. For convenience, hcp Co (110) and (103) peaks marked as the coloured rectangular bar, are considered for further discussion. These peaks are precisely fitted with the Gaussian function for different Co layer thicknesses, and zoomed portions of peaks for d=12 Å (island state) and 100 Å (continuous film state) are plotted in Fig. 4(b) and 4(c), respectively. As thickness increases to 100 Å, the two planes become distinctively separated, indicating an increase in crystallite size. Relative peak intensities may also be noted that in the island state (~12 Å), peak (103) is small as compared to peak (110). At higher thickness (~100 Å), the relatively (103) peak increases significantly as compared to the (110) peak. This is mainly due to the structural texturing, which might have appeared due to the internal stress developed during the film deposition [46]. It is known that crystallites nucleate with different crystallographic orientations during polycrystalline film growth. The grain boundaries are formed during this process when the islands grow to impinge and coalesce [45]. Due to the excess free energy at these boundaries, stress develops during the coalescence of islands [47]. At the different thicknesses during growth, the development of the stress is also associated with the island coalescence process [48] and is responsible for the texturing in the samples.

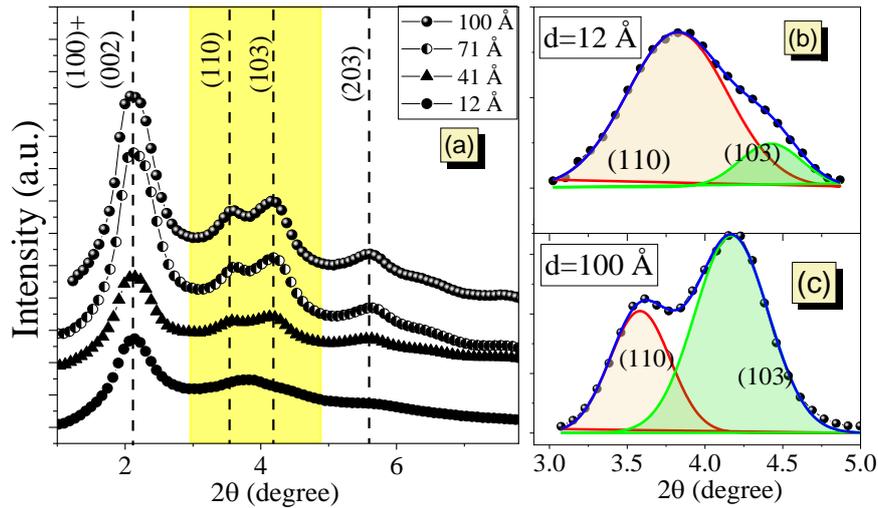

**Fig. 4.** (a) Radially averaged intensity from the RHEED images for the Co film at different thicknesses. (b) zoomed portion of peaks for Co thickness d=12 Å. (c) zoomed portion of peaks for Co thickness d=100 Å.

### 3.2.3 In-situ MOKE measurements

To study magnetic properties with Co film thickness, in-situ MOKE measurements were also performed simultaneously with the transport measurements in the same UHV chamber. Hysteresis loops were collected starting from the fraction of nanometer to study the $C_{60}$-Co interface precisely. Some representative MOKE hysteresis loops taken at the different thicknesses are shown in Fig. 5(a). This Figure shows no proper hysteresis loop was observed below 26 Å Co thickness. An appropriate MOKE loop was first detected after 26 Å film thickness deposition. The loop clearly shows the hysteresis loop with the finite coercive field at 26-27 Å deposition of Co film. The lack of hysteresis at the initial thickness indicates that the ferromagnetic phase has not been developed even after the deposition of 26 Å thick film, the loop with finite coercivity indicating the onset of ferromagnetism in the Co film. Further growth of Co film shows systematic variation in the coercivity and Kerr signal (height of the loop) with increasing thickness.

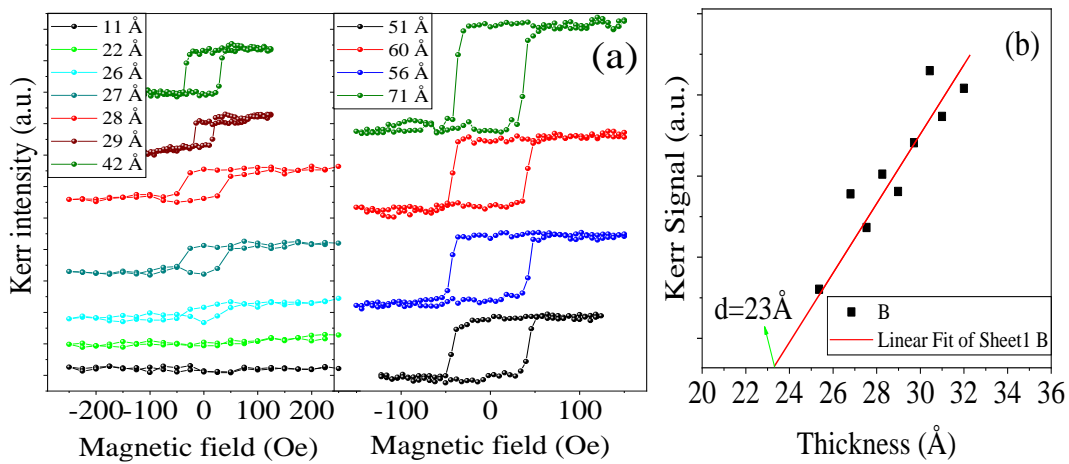

**Fig. 5.** (a) Some representative MOKE hysteresis loops of Co film deposited on $C_{60}$ (b) Variation in Kerr signal (Ms1-Ms2) as a function of the thickness of Co film.

One may note that the Kerr signal obtained from in-situ MOKE becomes a convenient tool to study the magnetic dead layer at the interface, which could be formed due to the possible intermixing, hybridization or some structural adaptations at the interfaces of inorganic layers [46,49]. The present case study of interface magnetism at the Co/$C_{60}$ (metal/organic layers) is of current interest, as they play a significant role in achieving reasonable magneto-transport in organic spin valves for practical applications. To get information about the dead layer at the interfaces, the thickness dependence of the Kerr signal (Ms1-Ms2) is extracted from the saturation point of hysteresis loops and is reported in Fig. 5(b).

The penetration depth of the laser beam is finite; thus, the strength of the Kerr signal ($\Phi_{long}$) in longitudinal geometry is expected to be a linear function of the thickness of the film and can be related to the expression as given below

$$\Phi_{long} = 4\pi n_{film}^2 Q\, d\theta / \lambda (n_{sub}^2 - 1)$$

where Q is the magneto-optical constant, d is the thickness of the magnetic film, θ is the angle of incidence of the laser beam from the surface normal, $n_{sub}$ and $n_{film}$ are the refractive index of the substrate and the film, respectively [50–52]. Thus, the signal is linearly proportional to the thickness in the ultra-thin film region and tends to saturate in a higher thickness range due to the limited penetration depth of laser [49,53].

One can see linearly fitted Kerr signal data in Fig. 5(b) that the signal is linearly proportional to the thickness. The extrapolated line cuts the x-axis at 23 Å thickness, suggesting the formation of the magnetic dead layer or magnetically inactive layer at the Co-$C_{60}$ interface. The formation of this ill-defined or magnetically inactive layer may be attributed to i) the formation of intermixed nonmagnetic phase or ii) at the interface, there are still some small islands and clusters, which are small enough and exhibit nonmagnetic or paramagnetic behavior. Therefore, do not contribute to the Kerr signal. It may also be noted that the dissociation of Carbon atoms from $C_{60}$ molecule is observed either after annealing at 900°C or under the bombardment of sputtering ions during deposition [54,55]. Therefore, the possibility of a nonmagnetic intermixed phase at the interface is negligible at room temperature.

### 3.3 Magnetic Anistropy Investigations of Co/$C_{60}$ bilayers

#### 3.3.1 X-ray reflectivity measurements

Fig. 6 (a)-(d) shows the X-ray reflectivity (XRR) patterns of all samples, which are plotted against the scattering vector $q_z = 4\pi \sin\theta/\lambda$ on the x-axis, where θ is the incident angle, and λ is the wavelength of the X-ray. The periodic oscillations (Kiessig fringes) in all the XRR patterns are due to the thickness of the total structure.[56] In the case of the S4 sample, broad and short periodic oscillations in the XRR pattern correspond to the Co and total thickness (Co+$C_{60}$), respectively. The difference between two periodic oscillations is inversely proportional to the thickness of the layer ($d = 2\pi/\Delta q$), where d is the film thickness. The broad oscillations are not clearly visible in the case of the thin Co layers. The more damping in the oscillations at the lower thickness of Co (sample S1, S2, and S3) is attributed to the rough interface due to interface mixing or deep penetration of Co metal atoms into the $C_{60}$ layer.[57]

| Sample | $Co_{top}$ | | $Co_{int}$ | |
|---|---|---|---|---|
| | d (Å) | ρ(g/cm³) | d (Å) | ρ (g/cm³) |
| S1 | - | - | 28 | 4.54 |
| S2 | 20 | 6.63 | 24 | 3.56 |
| S3 | 38 | 6.74 | 37 | 3.93 |
| S4 | 188 | 8.9 | 32 | 6.41 |

All XRR patterns are fitted using Parratt's formalism[58] to extract the structural parameters (layer thickness and interface mixing) to correlate the same with the evolution of magnetic properties at different thicknesses of the Co layer on $C_{60}$ film.

**Table 1** Thickness and density of Co film obtained from XRR fitting. Errors in layer thicknesses are ± 1 Å.

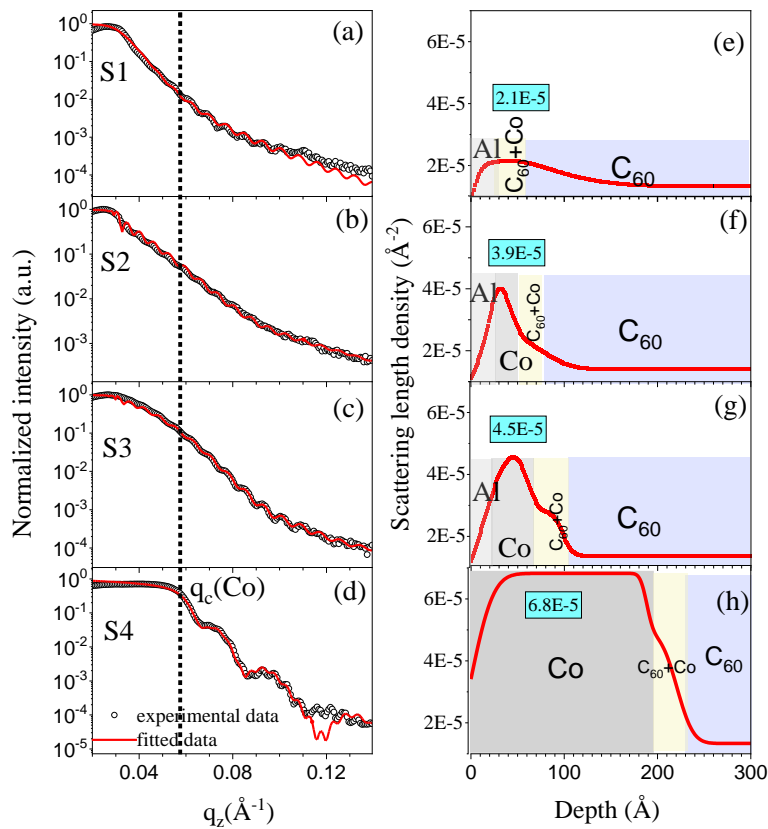

**Fig. 6** (a)-(d) X-ray reflectivity pattern, circles represent the experimental data obtained and solid line represents the fitted data and (e)-(h) scattering length density profile of the samples S1, S2, S3 and S4, respectively, where Co, $C_{60}$ and Al are used to represent the Cobalt, fullerene and Aluminium layer in the samples.

The best fit to the data is obtained by dividing the total deposited Co layer into two layers; *i)* interface Co layer (designated as $Co_{int}$) and *ii)* top Co layer (designated as $Co_{top}$) by considering two different electron densities in both layers. The corresponding Scattering length density (SLD) profiles, which measure the scattering power of a material and give the variation in the density of the material along the sample depth, are extracted and shown in front of the respective XRR pattern in Fig. 1(e)-(h). However, the SLD is related to the density because $\delta$ depends on the density ($\rho$) of the material. Thus, density ($\rho$) in g/cm$^3$ is calculated using extracted SLD along the depth and presented in Table 1 for $Co_{int}$ and $Co_{top}$ layers. $Co_{int}$ layers have lower $\rho$ in all samples due to Co diffusion in the $C_{60}$ layer near the interface. In the case of the S1 sample, XRR was fitted by taking the whole Co layer as the $Co_{int}$ layer. It is mainly due to the lower deposited thickness of Co film, which gets completely diffused inside the $C_{60}$ layer. The reduced electron density of the $Co_{top}$ layer (about 74% and 76% for samples S2 and S3) compared to bulk Co material ($Co_{bulk}$~8.9 g/cm$^3$) indicates that the $Co_{top}$ layer is still not entirely continuous.

On the other hand, the density of the $Co_{top}$ layer for S4 is almost the same as the Co bulk material, confirming the formation of a continuous layer on the $C_{60}$ film. This is per the *in situ* work, where Co film grows on $C_{60}$ *via* a diffused small isolated island in the $C_{60}$ matrix at the interface. These isolated islands grow larger to connect with other islands, which combine to form a continuous film at a higher thickness, around 75 Å.

### 3.3.2 MOKE measurements

The MOKE hysteresis loops of all the samples are shown in Fig. 2. No hysteresis loop is observed at lower thickness, i.e., in sample S1 (Fig. 7a). As the thickness of the Co layer increases in S2, the hysteresis loop starts to manifest almost without coercivity ($H_C$). However, in the case of samples S3 and S4, a well-defined $H_C$ of the loops can be seen in Fig. 7(c) & (d). These two samples have shown substantial variation in the shape of hysteresis loops with azimuthal angles. This drastic change in the shape of the hysteresis loops indicates the presence of UMA. The increasing curvature of the hysteresis with the azimuthal angle indicates a more significant role of magnetization rotation in the magnetization reversal process. In sample S4, the hysteresis curve along an azimuthal angle ($\phi = 0°$) is square-shaped, indicating that the domain wall motion dominates the magnetization reversal process.

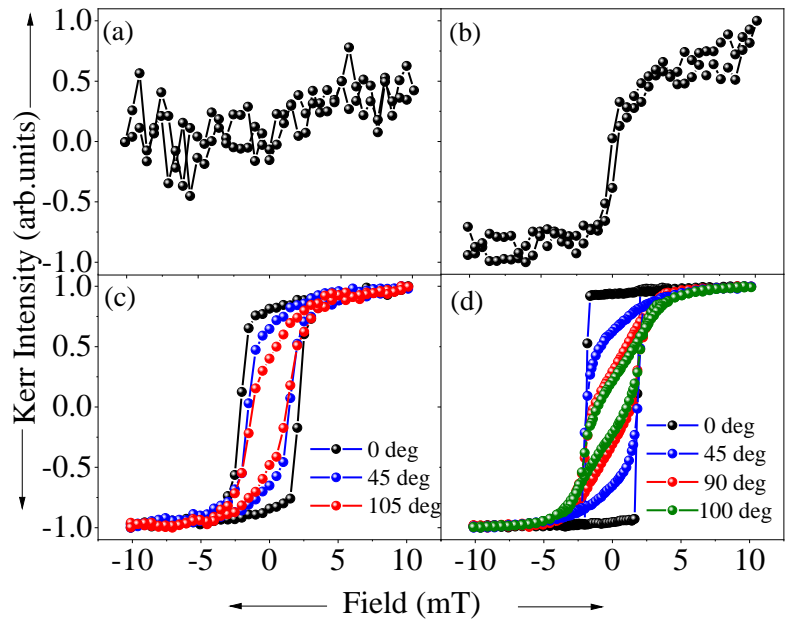

**Fig. 7** (a)-(d) Hysteresis loops along different magnetization directions for samples S1 -S4, respectively.

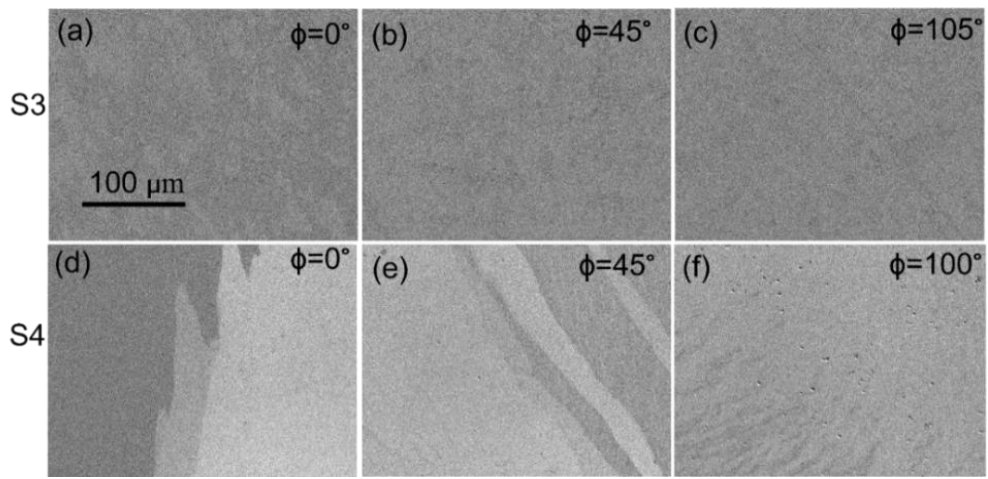

**Fig. 8** (a-c) and (d-f) shows the domain images recorded near coercive field values for different azimuthal angles $\phi$ for sample S3 and S4, respectively.

This is further confirmed by the Kerr microscopy domain images as depicted in Fig. 8(a)-(f). Here, the domain images at an azimuthal angle ($\phi = 100°$) (Fig. 8f) show ripple pattern domain, indicating that the magnetization rotation is incoherent.[59] It may also be noted that at some azimuthal angles, the Kerr images consist of three different shades, as shown in Fig. 8 (d & e), which suggests the presence of 2 different anisotropy axes. The large $H_C$ in sample S3 may be attributed to domain wall pinning, as seen in Fig. 8(a), which is taking place due to the interface roughness.[60]

The angular dependence of the $H_C$ is plotted in Fig. 9a and 9b, respectively, to understand the origin of the UMA in samples S3 and S4. The polar plots of $H_C$ variation show a kind of anomalous behaviour which neither corresponds to uniaxial nor biaxial magnetic anisotropy. In Co thin films grown on Si or quartz substrate, it is always observed to have a dumbbell-like structure in the $H_C$ variation with azimuthal angle, corresponding to the uniaxial magnetic anisotropy.[61–63]

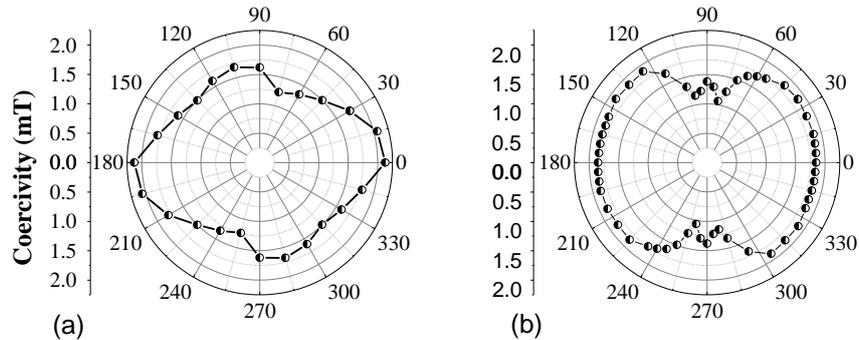

**Fig. 9** Azimuthal angle dependence of (a)-(b) coercivity (polar plots) for samples S3 and S4, respectively.

### 3.3.2 AFM Measurements

To correlate the observed magnetic properties with morphological and structural, samples were characterized using AFM, and XRD measurements. Fig. 10(a)-(d) gives the typical AFM micrographs for all samples. The average surface roughness values are obtained by extracting the line profiles from different parts of the AFM image. The final r.m.s. roughness values obtained are 22.4 Å, 20.5 Å, 26.1 Å and 22.0 Å for samples S1, S2, S3 and S4, respectively. The maximum roughness is obtained in sample S3. The size of the Co islands is found to increase systematically with increasing thickness of the Co layer on $C_{60}$. The mean size of Co grains is estimated using the ImageJ software and is obtained as 150 Å, 270 Å and 360 Å for samples S2, S3, and S4, respectively. Slight elongation of the particles observed along the marked arrow of the samples S3 and S4 images may be due to anisotropy in nucleation and growth or preferential percolation similar to the earlier study.[64]

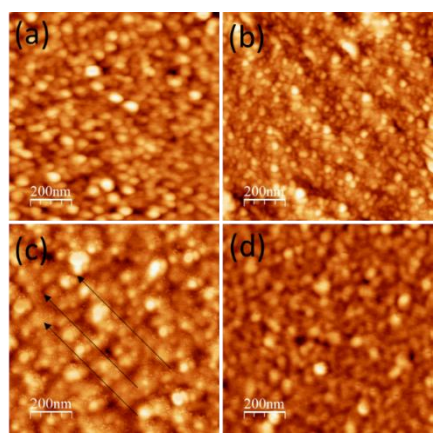

**Fig. 10** (a)-(d) shows AFM micrograph for sample S1, S2, S3 and S4 respectively.

### 3.3.3 X-ray diffraction measurements

To determine the possibility of any stress and its role in the origin of the UMA, the XRD of the higher thickness sample, i.e., S4, is performed in two geometries by keeping the momentum transfer vector ($q \sim q_{in}$ and $q_{out}$) in the plane (IP-XRD) and out (OP-XRD) of the film surface. Both geometries are shown in the schematic Fig. (9a and 9b). In OP-XRD geometry, the incident beam falls at a shallow glancing angle ($\theta_{in} \sim 1.5°$) on the sample, while the detector was scanned normal to the film surface. In this case, the $q_{out}$ is mainly in the film normal, and therefore, information is obtained about the scattering planes almost parallel to the film surface. In IP-XRD geometry, incident and diffracted beams make a grazing angle of 1.5° in the vertical direction, and the detector scans in the film plane. In this geometry, the momentum transfer vector ($q_{in}$) lies almost in the film plane; therefore, information is obtained preferentially from the scattering planes perpendicular to the film surface.

Fitted XRD patterns in IP and OP geometries are shown in Fig. 11 (c and d). The OP-XRD pattern of the S4 film exhibits three broad overlapping peaks, which correspond to (100), (002) and (101) planes of HCP-Co. On the other hand, in IP-XRD, mainly the HCP (002) peak appeared, suggesting oriented HCP (002) growth in the film plane.

The line widths for the HCP (002) peak (0.07 Å$^{-1}$) in OP-XRD patterns are broader as compared to those in IP-XRD (0.04 Å$^{-1}$). This suggests that grains are elongated in the film plane. To get qualitative information, position ($q$), normalized relative intensity ($I_A$), and d-spacings ($d$), corresponding to HCP (100), (002), and (101), are extracted after fitting the obtained XRD patterns and shown in Table 2. It is important to note that the d-spacings corresponding to all three planes, namely (100), (002), and (101), have reduced in IP-XRD with respect to OP-XRD, suggesting in-plane compressive stress in Co film in the S4 sample. One can also notice that along with the stress, in-plane texturing along hcp (002) takes place in the Co film.

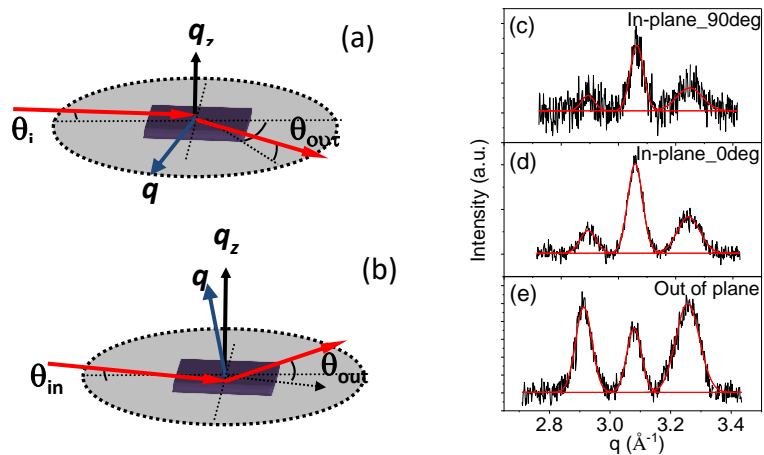

**Fig. 11.** (a) and (b) Geometries used for the IP-XRD and OP-XRD measurements, where q is the momentum transfer vector in the film plane ($q_{in}$) and normal to the film plane ($q_{out}$), respectively. (c) and (d) Fitted XRD patterns in IP-XRD geometries, (e) OP-XRD geometry.

**Table 2.** Fitting parameters obtained from XRD patterns taken along in-plane (IP) and out-of-plane (OP) geometry for S4; typical error bars obtained from the least square fitting of the XRD data approx. ± 2% in the value of the normalized area, $I_A$.

| Geometry | $q$ (Å$^{-1}$) | $I_A$ (%) | $d$ (Å) | (hkl) |
|---|---|---|---|---|
| **OP-XRD** | 2.91 | 0.31 | 2.18 | (100) |
| | **3.08** | **0.21** | **2.06** | **(002)** |
| | 3.26 | 0.47 | 1.95 | (101) |
| **IP-XRD_0deg** | 2.89 | 0.14 | 2.16 | (100) |
| | **3.05** | **0.53** | **2.04** | **(002)** |
| | 3.24 | 0.32 | 1.91 | (101) |
| **IP-XRD_90deg** | 2.89 | 0.12 | 2.16 | (100) |
| | **3.06** | **0.52** | **2.04** | **(002)** |
| | 3.24 | 0.35 | 1.91 | (101) |

The combined analysis can be understood well with the model, as shown below in Fig. 12. At low-thickness, i.e., sample S1, the Co clusters emerge in island state formation with varying sizes, leading to the formation of Cobalt enriched layer of $C_{60}$ having thickness of around 28 Å. Due to the small size of Cobalt clusters, they do not exhibit hysteresis loop due to the absence of ferromagnetism. With increasing thickness, the island size increases in sample S2 and manifests hysteresis loops almost without coercivity due to the superparamagnetic behaviour of Co clusters.

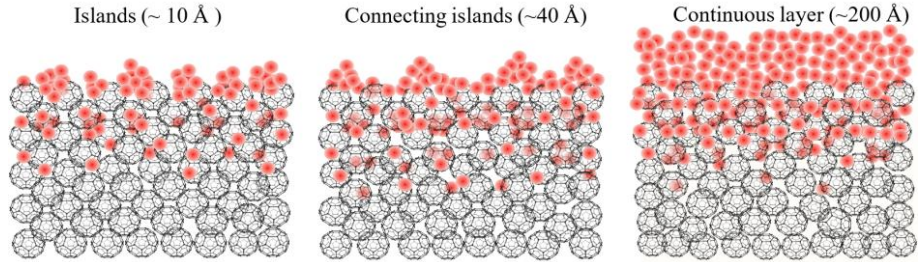

**Fig. 12.** Schematic of the growth evolution of the Co atoms (red balls) on the $C_{60}$ layer (black buckyballs).

As the surface free energy of Co is substantially higher than $C_{60}$ surface energy,[65,66] Volmer–Weber type of growth is expected in the present case, where islands of Co nucleate in and on the $C_{60}$ layer. The same has also been observed from resistivity measurements,[67] where the conduction of electrons from island to island does not take place up to 14 Å coverage due to the separation between Co islands. With further increasing thickness of the Co layer, i.e., above 20 Å, the formation of the percolating path between the isolated islands takes place, and the film becomes continuous after deposition at 75 Å thickness. From the structural characterizations, i.e., XRD measurements, in-plane texturing of Co film as well as compressive stress in the film plane, is confirmed. This texturing of the Co film along hcp (002) is the major cause of magneto-crystalline anisotropy in our sample, while the stress and the preferential orientation of Co grains are giving rise to magnetostrictive.

The unusual observed in the angular dependence of coercivity is explained in terms of the presence of dispersion in the magnetic anisotropy. This was further confirmed by the domain images taken in transverse geometry. Fine ripple magnetic domains were observed in the hard axis of magnetization in sample S4. The formation of ripple domains, i.e., magnetic fluctuations, is caused by random distribution of magnetocrystalline anisotropy, the inhomogeneous strain, or the presence of non-magnetic impurities.[68–70] It is shown in the literature that it is the misalignment in the local magnetic anisotropy leads to coercivity and remanence larger than zero near hard axis. In contrast to the UMA in Co/Si substrate, where well-defined UMA developed with one easy and hard magnetization axis, Co/$C_{60}$ bilayer gives two UMA components with unusual angular dependence of the coercivity. The origin of such magnetism can be understood in terms of the modified growth of the Co on the $C_{60}$ layer due to the soft nature of the underneath $C_{60}$ layer.

### 3.4 Study of Co₂FeAl (CFA)wedge in the proximity of $C_{60}$

#### 3.4.1 X-ray refelctivity characterizations

XRR patterns are taken at different positions, d1 to d5, along the length of the sample. It can be seen in Fig. 13, that there are dips below the critical angle of Pd, which increases with increasing thickness of the CFA layer. These dips correspond to the formation of X-ray standing wave inside the $C_{60}$ cavity. The structural parameters of the wedge sample are determined by the X-ray reflectivity data fitted with the Parratt formalism. The obtained experimental data points and the fitted curve in red reproduce well the intensity modulation. Fig. 13(b) shows the electron scattering length profile at respective positions in the wedged sample. It is to be noted that the XRR data is best fitted by considering two different layers of CFA, denoted as $CFA_{int}$ and $CFA_{bulk}$. Due to the diffusion of CFA clusters inside the $C_{60}$, an interface layer denoted as $CFA_{int}$ is formed with a comparatively lower density than $CFA_{bulk}$. It is observed that with the increasing thickness of the CFA, the density of the $CFA_{int}$ layer also increases. The deposited thickness of the $CFA_{bulk}$ layer at positions d1, d2, and d3 is found to be 97.8 Å, 63.8 Å, and 30.8 Å. The obtained thickness gradient of the $CFA_{wedge}$ is 2 Å/mm, and the thickness of the remaining layers remains almost similar throughout the wedge. The values obtained for the thickness, roughness, and density of the layers are mentioned in Table 3.

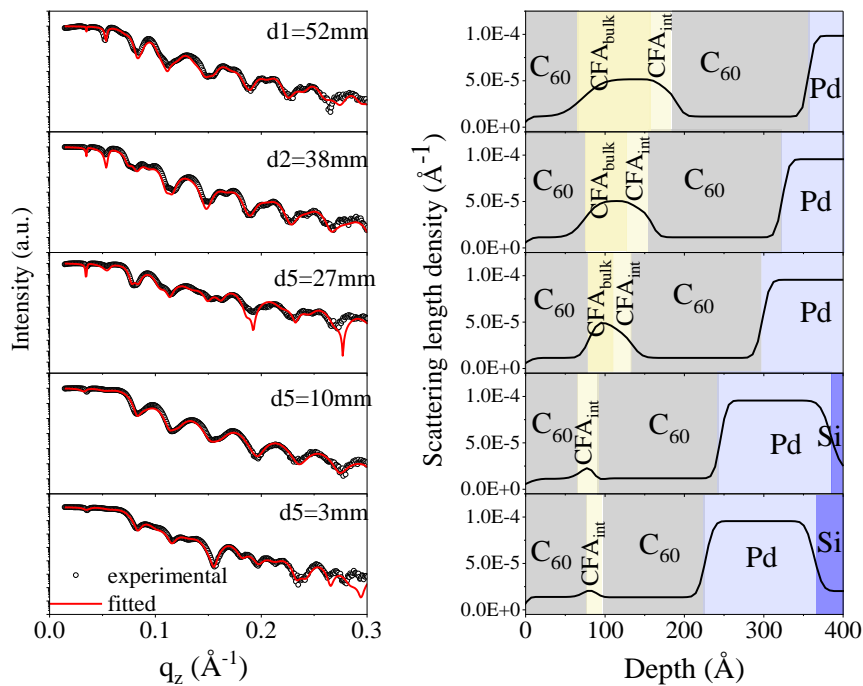

**Fig.** 13 (a) & (b) X-ray reflectivity pattern and Scattering length density profile of the sample structure (Si/Pd/$C_{60}$/ $CFA_{wedge}$/$C_{60}$) at different positions along the length.

**Table 3** Thickness, roughness, and density of all the layers in the sample obtained from XRR fitting. Errors in layer thicknesses and roughness are ± 1 Å.

| Position (mm) | | Pd | $C_{60}$ | $Co_2FeAl$ (diff) | $Co_2FeAl$ (bulk) | $C_{60}$ |
|---|---|---|---|---|---|---|
| 52 | Thickness (Å) | 145.5 | 165.3 | 21 | 97.8 | 72.1 |
| | Roughness(Å) | 6 | 7.5 | 8 | 20 | 5.2 |
| | density (g cm$^{-3}$) | 12 | 1.34 | 5.37 | 6.98 | 1.34 |
| 38 | Thickness | 144.3 | 164.1 | 21.6 | 63.8 | 73.9 |
| | Roughness | 6.5 | 6.9 | 8.6 | 13.8 | 6.8 |
| | density (g cm$^{-3}$) | 12 | 1.34 | 5.31 | 6.85 | 1.34 |
| 27 | Thickness | 141.6 | 164 | 24.1 | 30.8 | 79.5 |
| | Roughness | 7.2 | 9.7 | 9.1 | 8.2 | 8.1 |
| | density (g cm$^{-3}$) | 12 | 1.32 | 5.31 | 6.85 | 1.34 |
| 10 | Thickness | 139 | 160.3 | 10 | | 75.8 |
| | Roughness | 6.6 | 5.5 | 13 | | 11.1 |
| | density (g cm$^{-3}$) | 12 | 1.38 | 4.81 | | 1.32 |
| 3 | Thickness | 136.5 | 139.7 | 8 | | 79 |
| | Roughness | 7.8 | 7.1 | 10 | | 4.3 |
| | density (g cm$^{-3}$) | 12 | 1.58 | 4.38 | | 1.63 |

## 3.4.2 Thickness dependent MOKE measurements

To investigate the magnetic properties of the sample, hysteresis loops are taken in longitudinal geometry, at different positions across the sample with increasing thickness of CFA. Figure 14(a) and (b) shows the schematic of the sample structure and obtained hysteresis loops taken at positions marked as A1 to A4, respectively. At the bottom of the wedge ,i.e., at position A1, no hysteresis is observed. With increasing thickness of the CFA, hysteresis loop with a low coercive field begins to appear at a thickness of 28 Å i.e., at A2. A well-defined square-shaped hysteresis loop having coercivity of 1.2 mT is observed at position A3, i.e., in the mid of the wedge. The shape of the hysteresis loop changes with the increasing thickness of the wedge as can be seen at position A4, i.e., at the top. The variation in the shape of the hysteresis loop suggests the rotation of the magnetic anisotropy axis within the sample. As Continuous loops are taken along the length, i.e., towards the increasing thickness of the CFA$_{wedge}$, a systematic variation in the Kerr signal as well as coercive field is found to be observed as shown in Fig. 15(a) & (b) .

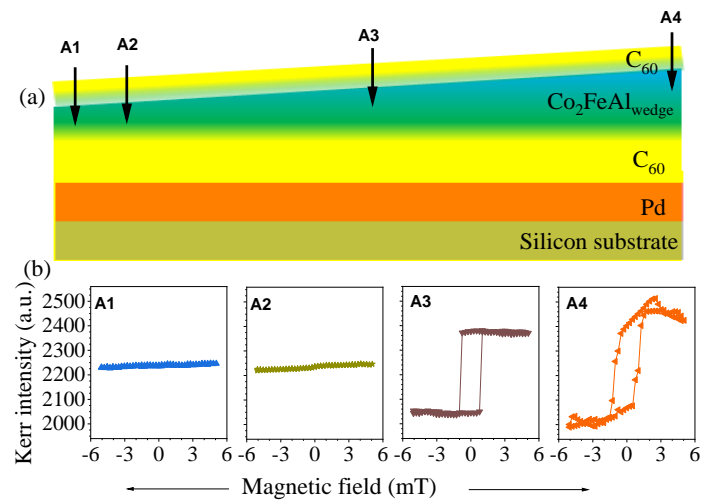

**Fig.** 14 (a) Representative hysteresis loops along the length of the sample

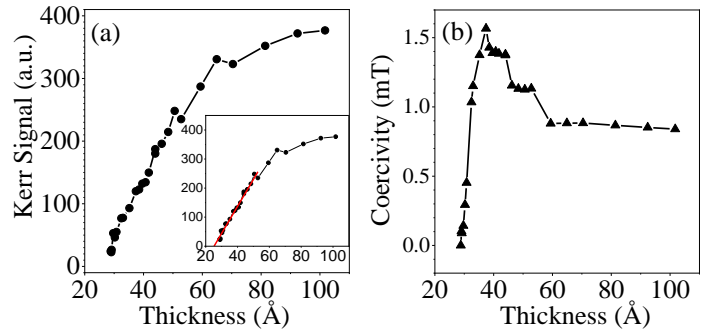

**Fig.** 15 (a) & (b) shows the variation of the Kerr Signal and coercivity w.r.t. thickness of $Co_2FeAl_{wedge}$

The Inset of Fig. 15(a) shows the linearly fitted Kerr signal data which suggests that Kerr signal increases in proportion with the thickness of the wedge. The extrapolated line cuts the horizontal x-axis at 25 Å thickness, forming a magnetically inactive layer of $Co_2FeAl$ showing no ferromagnetic behaviour. The coercivity vs. thickness plot in Fig. 15(a) shows that coercivity increases initially with increasing thickness, and after reaching the maximum value, it starts decreasing and then reaches saturation. The variation in the shape of the hysteresis loop with the increasing thickness is understood precisely by performing MOKE measurements at different positions w.r.t. azimuthal direction. Fig.16 (a-f) represents the hysteresis loops taken at different positions of the sample corresponding to different thicknesses of the $Co_2FeAl$ along with their polar plots of remanent magnetization ($M_R/M_S$) and coercive field ($H_C$). The depicted hysteresis loops correspond to the magnetic easy axis, hard axis, and the one with the highest coercive field. The variation in the shape of the hysteresis in different azimuthal directions suggests the presence of magnetic anisotropy in the film plane.

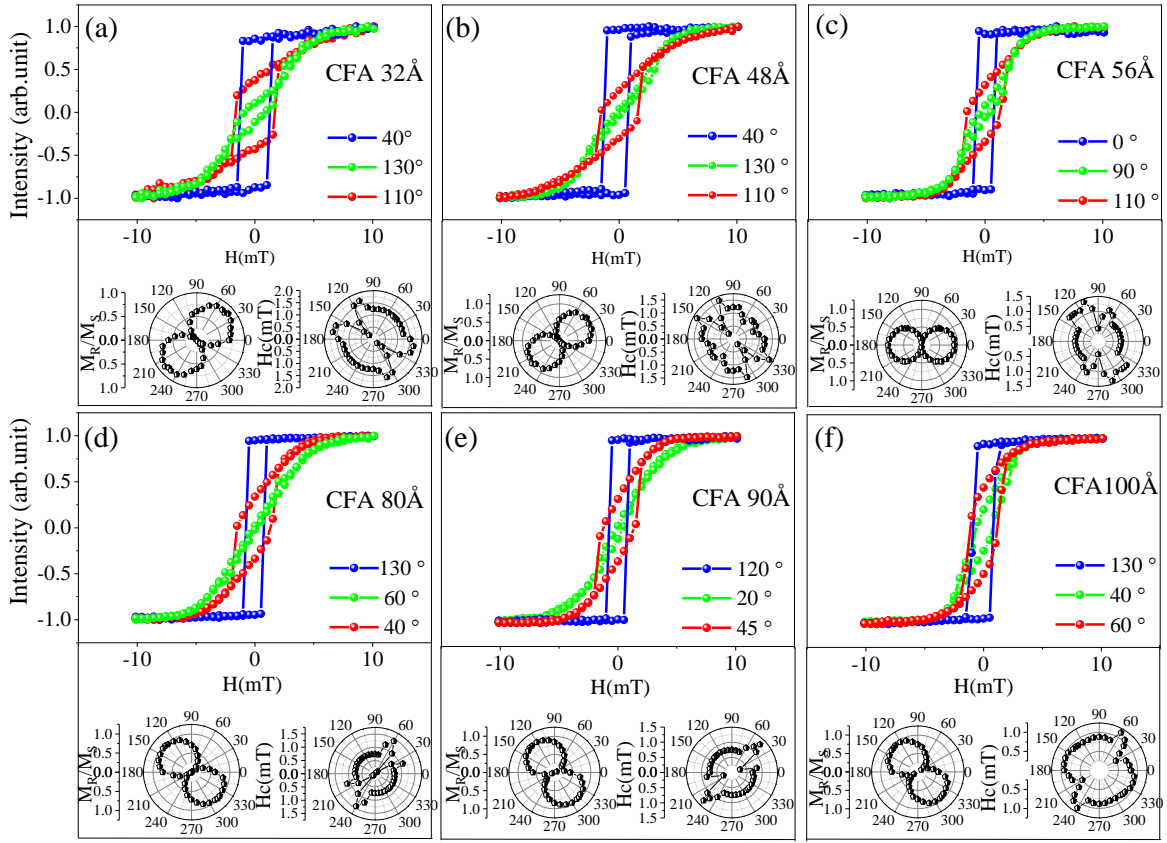

**Fig. 16** Representative hysteresis loops in different azimuthal angle at different thicknesses of $CFA_{wedge}$ and the corresponding polar plots of coercivity and remanent magnetization.

The remanence variation w.r.t. azimuthal angle further confirms the presence of two-fold uniaxial magnetic anisotropy. The direction of the magnetic easy axis varies with different thickness positions of the wedge. However, the preparation conditions, such as preparation technique and angle of incidence and underneath layer, are identical throughout the wedge.

The azimuthal angle dependence of remanence can be modeled to more precisely determine the direction of the magnetic easy axes. The projection of the in-plane magnetization vector to the plane of incidence of light is a cosine-like function [33] thus, the two-fold symmetry of the magnetic anisotropy is fitted by,

$$\frac{M_R}{M_S} = \frac{M_R^{max}}{M_S} |\cos(\phi - \phi_0)| + \frac{M_R^{off}}{M_S}$$

Where, $M_R^{max}/M_s$ is the strength of the Uniaxial magnetic anisotropy, $M_R^{off}/M_s$ is the isotropic contribution, and $\phi_0$ is the phase or the direction of the magnetic easy axis. After determining the direction of the magnetic easy axis at different positions using this analysis technique, one can easily understand the rotation of the magnetic anisotropy axis with CFA thickness. The angle of the magnetic axis is around 30° and then changes to 0° in the middle of the wedge and then changes to 120° and remains constant till the maximum thickness of the wedge has been achieved. The change in the strength of magnetic anisotropy and its direction may be attributed to the variation of the internal stress and structure texture along the length of the wedge (with increasing thickness of CFA).

## 4. Conclusions

In conclusion, the growth of the ferromagnetic thin film on $C_{60}$ has been studied in detail. The in-situ growth of Co film on $C_{60}$ has been studied in-situ using MOKE, RHEED and Transport. The 23-25 Å thick magnetic dead layer at the metal-$C_{60}$ interface is observed which is formed due to the formation of small clusters with superparamagnetic properties. Further, analysis reveals that the growth of Co film takes place via island formation and combines into a continuous film at a thickness of around 75 Å, which is much larger compared to the Co film on Si. Structural texturing with increasing thickness of the Co film is attributed to the formation of internal stresses that might have been generated during the coalescence of Co islands. The origin of the unusual magnetic anisotropy in Co/$C_{60}$ is attributed to the misalignment in the local magnetic anisotropy. The change in the strength of magnetic anisotropy and its direction may be attributed to the variation of the internal stress and texture with increasing thickness of CFA. The interface between the top electrode and organic layer in the case of OSVs is still a matter of deep understanding due to the diffusion of top ferromagnetic electrodes taking place in the organic layer.

## Acknowledgement

Arun Singh Dev and Md. Shahid Jamal is acknowledged for their help in In-situ measurements and constructive discussions. Er. Ajay Rathore and Dr. V.G. Sathe are thanked for Raman measurements. Thanks to Mr. Mohan Gangrade for the AFM measurements. I also thank Dr. Pooja Gupta and Dr. Sanjay K. Rai for synchrotron based XRD measurements at RRCAT, Indore. Author also acknowledge Dr. Mukul Gupta and Ms. Akshaya for Cobalt deposition on top of $C_{60}$ layer.